\documentclass[10pt, letterpaper]{article}

\usepackage{algorithm}
\usepackage{algorithmic}
\usepackage{amsmath}
\usepackage{amsfonts}
\usepackage{amssymb}
\usepackage{amsthm}
\usepackage{caption}
\usepackage{color}
\usepackage{enumerate}
\usepackage{epsfig}
\usepackage[left=3cm, right=3cm, top=3cm, bottom=3cm]{geometry}
\usepackage{graphicx}
\usepackage{mathrsfs}
\usepackage{multicol}
\usepackage{txfonts}

\DeclareCaptionLabelFormat{graphs}{Exhibit~#2}

\newcommand{\rnVar}{\widetilde{Var}}

\newcommand{\rnE}{\widetilde{\textup{E}}}

\newcommand{\E}{\textup{E}}
\newcommand{\Var}{\textup{Var}}

\title{A New Trinomial Recombination Tree Algorithm and Its Applications}
\author{Peter C.L. Lin \\ Johns Hopkins University \\ peter.lin@jhu.edu}

\begin{document}

\maketitle

\section{Introduction}

\begin{figure}[tp]
\captionsetup{labelformat=graphs}
\centering
\includegraphics[width=0.3\textwidth, angle=270]{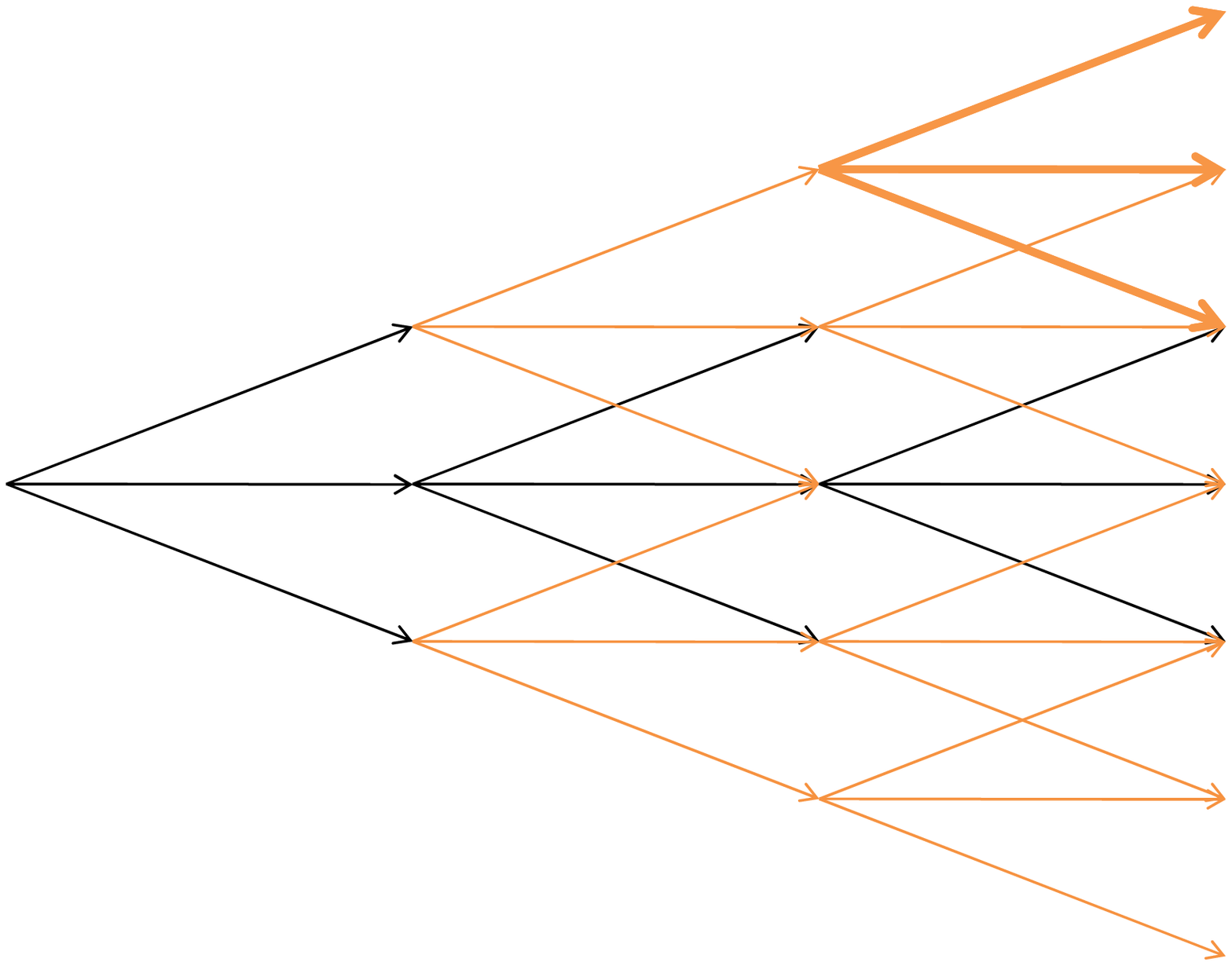}
\includegraphics[width=0.3\textwidth, angle=270]{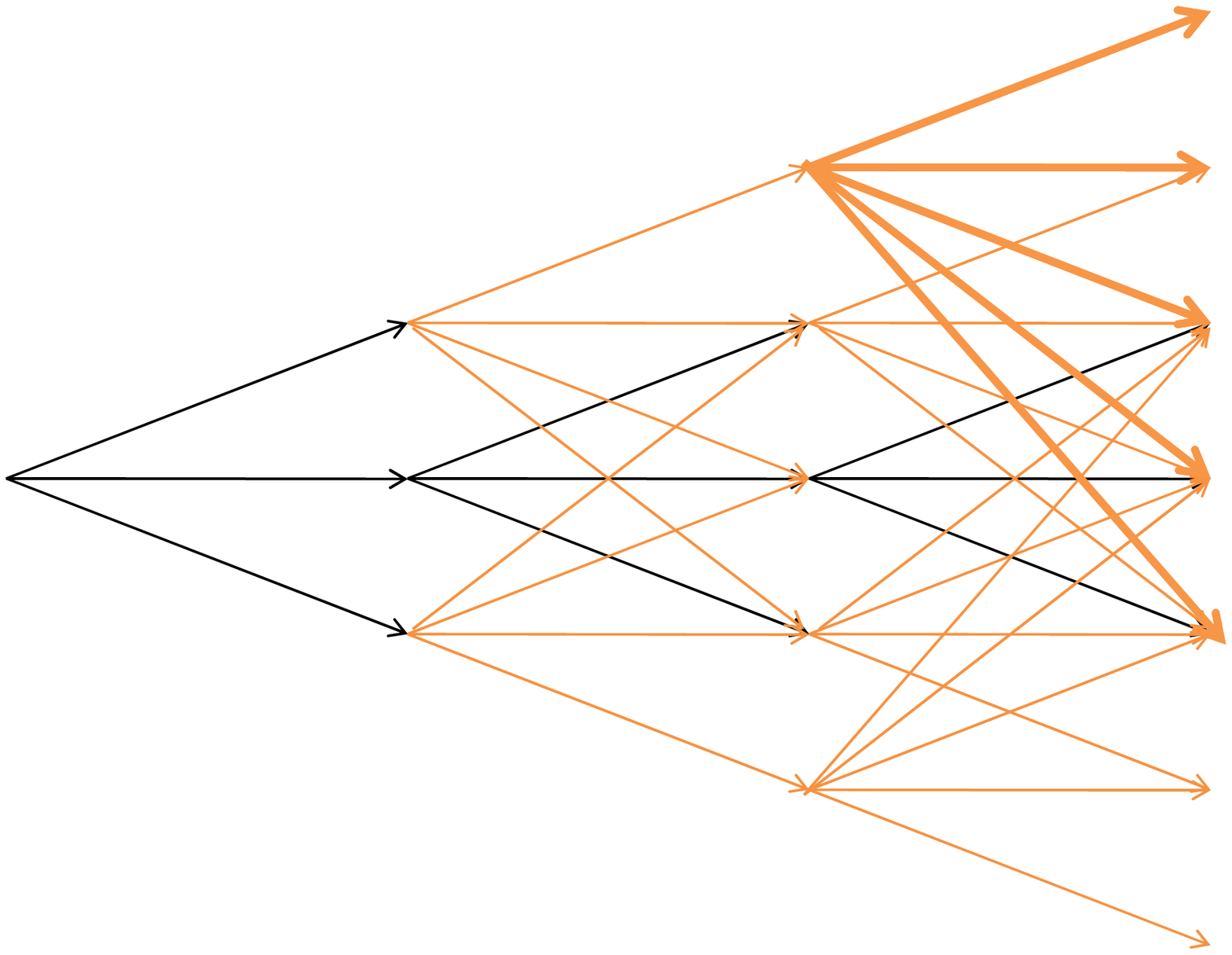}
\includegraphics[width=0.3\textwidth, angle=270]{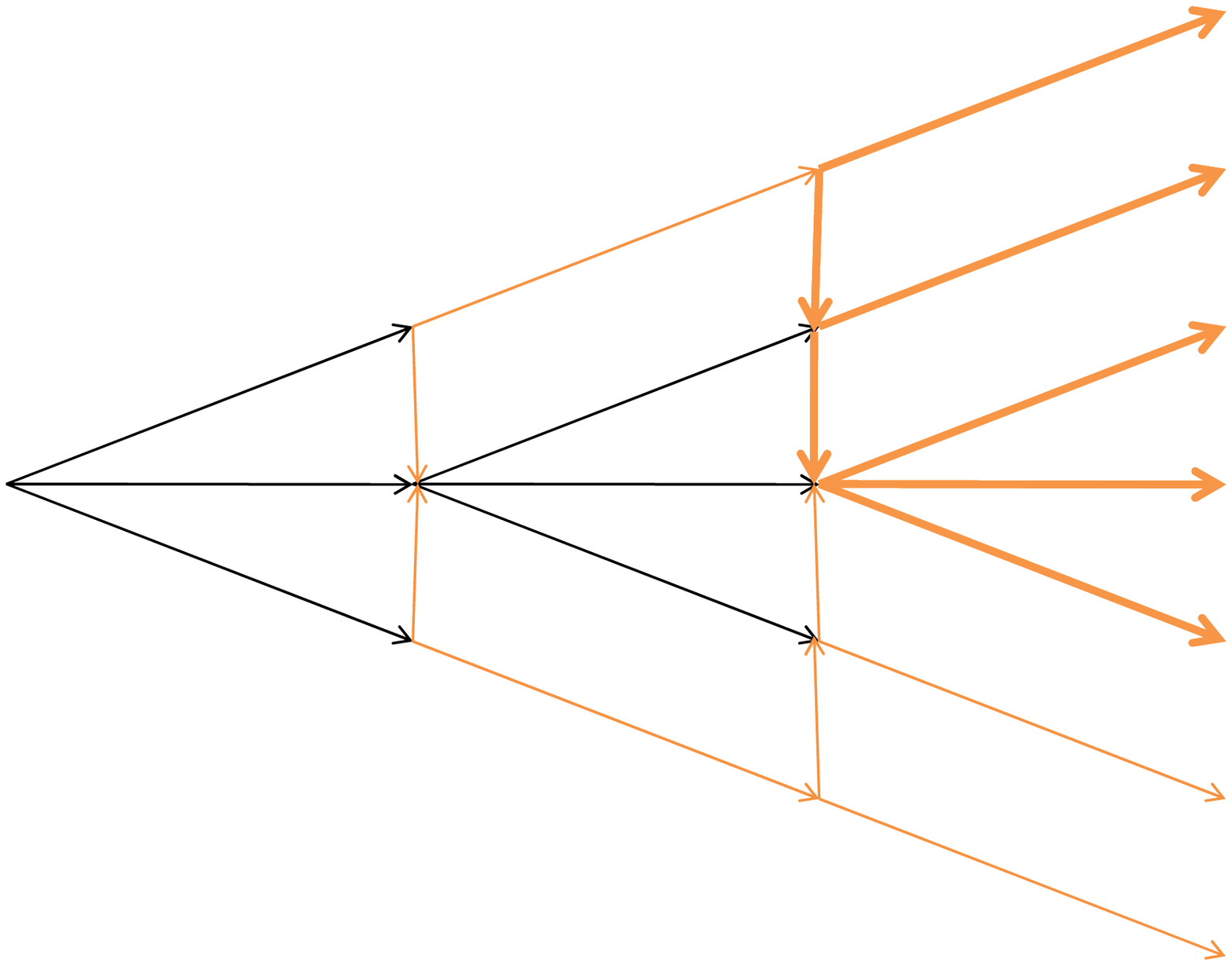}
\caption{[Top-Left] Original Recombination Trinominal Tree; [Top-Right] Modified Tree; ; [Bottom] Simplified Tree}
\label{ssm_tree}
\end{figure}


A tree is an acyclic structure where each node has zero to multiple descendent nodes and one parent node. A recombination tree is a special tree structure of which the size grows linearly. Therefore the investing decisions, if computed recursively, has time complexity\footnote{For the terminology of computational complexity please see Cormen \textit{et al.}, \cite{clrs}.} at most $O(n^2)$, which is much more efficient than a general simulation method which may cost exponential amount of time. For example, a recombination tree can help us to efficiently determine the price and the buy/sell timing for an American style option by comparing the derivatives value at each tree node with its children nodes. However, designing a recombination tree algorithm for modeling interest rates is far from trivial. Here are two examples: \\

Hull and White \cite{hw1990} provided a heuristic two-stage method for constructing an interest rate tree based on the extended-Vasicek short rate model. In the first stage the algorithm builds the framework of the tree, and in the second stage the algorithm calibrates the tree to the current interest-rate term structure. The algorithm is designed for a short-rate model; hence the tree cannot be adjusted or updated according to the markets. Also, their method cannot deal with stochastic mean-reverting parameters, and there is no guarantee that the tree is a recombination tree especially when the volatility term of the short-rate process is a decreasing function. Therefore, Hull-White's algorithm is not a good candidate.   \\

Black, Derman, and Toy \cite{bdt1990} (BDT hereafter) also provided a recombination tree algorithm for short rates. Their tree is constructed recursively and calibrated to zero-coupon bond volatilities and current interest rate structure. Though the BDT tree guarantees a recombination structure, the tree is not designed for a general Ornstein-Uhlenbeck process. Therefore, the BDT tree is not a good candidate either. \\

In light of the fact that no existing tree algorithms can guarantee the recombination property for general Ornstein-Uhlenbeck processes, we propose a new recombination trinominal tree algorithm for our stochastic-splines model (this is formalized below). The idea is to modify a standard trinominal tree (Exhibit \ref{ssm_tree} [Top Left]) by adding extra branches at each node. First, we denote the tree structure in black color the \textit{center path}. Then, given a node $V$ directly above (below) the center path, define $\langle V \rangle$ the set of nodes containing $V$ and all the nodes down to (up to) the center path. Then we modify the tree according to the following rules: (i) the center path remains unchanged; (ii) given a node $V$ above the center path, we connect the node to all the descendant (children) nodes stemming from $\langle V \rangle$; (iii) given a node $V$ below the center path, we connect the node to all the descendant nodes stemming from all the nodes from $\langle V \rangle$. The modified tree structure is shown in Exhibit \ref{ssm_tree} [Top Right]. We will use the names, \textit{spanning nodes} and \textit{spanning branches}, to identify those nodes not on the center path, and branches not emanating from a center node. \\

The crucial key of modifying a standard trinominal tree is that we can further simplify and still keep the tree structure by adding \textit{sibling branches}. Sibling branches are one-way streets through which we can only move up or down at a given time epoch, but not in both directions. A spanning node above the center path can reach the center path and all the nodes in between only by moving down through the sibling branches; a spanning node below the center path can reach the center path and all the nodes in between only by moving up through the sibling branches. As a result, by adding the sibling branches, each node can reach all but one descendent nodes via its sibling branches. So, in the simplified tree structure, each spanning node will have only one time transition descendant branch. The final tree structure is shown in Exhibit \ref{ssm_tree} (Bottom). The algorithm is given below. The proof of the correctness of the algorithm for simulation is given in Section 3.  \\

\section{Algorithm}

Now we give a full description of the algorithm. Let $g_{j,j}$ denote the node on the center path at time $t_j$, and let $g(t_j,j)$ denote the value of node $g_{j,j}$. Therefore, if $g: \mathbb{R}\times \mathbb{Z}\mapsto \mathbb{R}$ is represented as a function, then it indicates the value of the node. Let $g_{j,j+k}$ and $g(t_j,j+k)$ denote the $k$-th node above center node $g_{j,j}$ and the value of $g_{j,j+k}$ respectively. Similarly, let $g_{j,j-k}$ and $g(t_j,j-k)$ denote the $k$-th node below center node $g_{j,j}$ and the value of $g_{j,j-k}$ respectively. Moreover, if we use capital letter $G(t_j, \omega)$, then it represents a random variable of the tree value at time $t_j$. To shorthand the notation, the expectation value $G(t_{j+1})$ conditional on the position of $G(t_j)$ is written as
\begin{align}
	\E\left[ G(t_{j+1}) \mid G(t_{j}) \right].
\end{align}
Define the conditional expectation at node $g_{j,j}$ to be
\begin{align}
	M(t_j,j) = \E\left[ G(t_{j+1}) \mid G(t_j) = g_{j,j} \right]
\end{align}
Since the volatility term in stochastic-splines model is assumed to be a deterministic function, the conditional variance is the same for all nodes at a given time, i.e. at time $t_j$, the conditional variance is
\begin{align}
	V_j = \Var\left( G(t_{j+1}) \mid G(t_j) \right).
\end{align} 

The idea of the recombination algorithm is to construct the center path first including the node values and branch probabilities, then determine the values of the spanning nodes, the probabilities on the spanning branches, and the probabilities on the sibling branches. The details are given in Algorithm \ref{tree_algo}. However, the algorithm shows that each tree is designed for simulating one coefficient process; if we have $N$ coefficient, we will need to build $N$ trees altogether if coefficient processes are correlated. After constructing the coefficient recombination "forest", we can simulate the interest rate curve efficiently. \\


\begin{algorithm}[tp]
\centering
\caption{Recombination Tree}
\begin{algorithmic}[1]
\REQUIRE \textbf{MAXLEVEL} (Tree Size)
\bigskip
\bigskip

\STATE\COMMENT{\textit{\textbf{Stage One - Center Path}}}
\bigskip
\FOR{$j=1$ to \textbf{MAXLEVEL}}	
	\STATE $\Delta x_j \leftarrow \sqrt{V_j}$
	\STATE $h \leftarrow \mbox{round}\left(\frac{M(t_j,j)}{\Delta x_j}\right)$
	\STATE Set
		$\left\{ \begin{array}{cl}
			g(t_j,j) & \leftarrow h \times \Delta x_j \\
			g(t_j,j+1) & \leftarrow (h+1) \times \Delta x_j \\
			g(t_j,j-1) & \leftarrow (h-1) \times \Delta x_j
		\end{array} \right..$
	\STATE Set
		$\left\{ \begin{array}{cl}
			p_u & \leftarrow \frac{1}{6} + \frac{(M(t_j,j) - g(t_j,j))^2}{6V_{t_j}^2} + \frac{M(t_j,j) - g(t_j,j)}{2\sqrt{3}V_{t_j}} \\
			p_n & \leftarrow \frac{2}{3} - \frac{(M(t_j,j) - g(t_j,j))^2}{3V_{t_j}^2} \\
			p_d & \leftarrow \frac{1}{6} + \frac{(M(t_j,j) - g(t_j,j))^2}{6V_{t_j}^2} - \frac{M(t_j,j) - g(t_j,j)}{2\sqrt{3}V_{t_j}}
		\end{array} \right..$
\ENDFOR
\bigskip

\STATE\COMMENT{\textit{\textbf{Stage Two - Spanning Branches}}}
\bigskip
\FOR{$i=1$ to \textbf{MAXLEVEL}}
	\FOR{$j=1$ to $2\times j -1$}
		\STATE\COMMENT{\textit{\textbf{Move to the $j$-th vertex below the center path}}}
		\STATE Find $x \in [M_{i,j-1}-\sqrt{V_i},M_{i,j-1}+\sqrt{V_i}]$ where
			\begin{align*}
				\frac{x - M_{i,j-1}}{x - M_{i,j}} = \frac{V_i - (x - M_{i,j-1})^2}{V_{j} + (M_{i,j} - M_{i,j-1})^2 - (x - M_{i,j-1})^2}
			\end{align*}
		\STATE $g(t_{i+1},j-2) \leftarrow x$
		\STATE $p(g_{i,j-1},g_{i,j}) \leftarrow \frac{x - M_{i,j-1}}{x - M_{i,j}}$
		\STATE $p(g_{i,j-1},g_{i+1,j-2}) \leftarrow 1-\frac{x - M_{i,j-1}}{x - M_{i,j}}$
		\medskip
		\STATE\COMMENT{\textit{\textbf{Move to the $j$-th vertex above the center path}}}
		\STATE Find $x \in [M_{i,j+1}-\sqrt{V_i},M_{i,j+1}+\sqrt{V_i}]$ where
			\begin{align*}
				\frac{x - M_{i,j+1}}{x - M_{i,j}} = \frac{V_i - (x - M_{i,j+1})^2}{V_{j} + (M_{i,j} - M_{i,j+1})^2 - (x - M_{i,j+1})^2}
			\end{align*}
		\STATE $g(t_{i+1},j+2) \leftarrow x$
		\STATE $p(g_{i,j+1},g_{i,j}) \leftarrow \frac{x - M_{i,j+1}}{x - M_{i,j}}$
		\STATE $p(g_{i,j+1},g_{i+1,j+2}) \leftarrow 1-\frac{x - M_{i,j+1}}{x - M_{i,j}}$
	\ENDFOR
\ENDFOR
\end{algorithmic}
\label{tree_algo}
\end{algorithm}

\section{Verification}

First we look at the \textit{first part} of the algorithm and some notations. The first stage of the algorithm follows the standard Hull-White methodology and provides the backbone of the tree. Let $g_{j,j}$ denote the node on the center path at time $t_j$, and let $g(t_j,j)$ denote the value of node $g_{j,j}$. Therefore, if $g$ is represented as a function, then it indicate the value of the node. Let $g_{j,j+k}$ and $g(t_j,j+k)$ denote the $k$-th node above center node $g_{j,j}$ and the value of $g_{j,j+k}$ respectively. Similarly, let $g_{j,j-k}$ and $g(t_j,j-k)$ denote the $k$-th node below center node $g_{j,j}$ and the value of $g_{j,j-k}$ respectively. Moreover, if we use capital letter $G(t_j)$, then it represents a random variable of the tree value at time $t_j$. To shorthand the notation, the expected value $G(t_{j+1})$ conditional on the position of $G(t_j)$ is written as
\begin{align}
	\E\left[ G(t_{j+1}) \mid G(t_{j}) \right].
\end{align}

Now we move to the \textit{second part} of the algorithm. The tree branches besides the central path are called spanning branches and spanning nodes. The second stage of the algorithm adopts the ideas of the law of total expectations and the law of total variances to assign the values and probabilities of spanning nodes and branches. The procedure is done recursively. Therefore we just need to look at the cases when $k = -1$ and $k = 1$. Given a node $g_{j+1,j+2}$ spanning from node $g_{j,j+1}$ at time $t_j$, we denote the conditional expectation and conditional variance at node $g_{j,j+1}$ to be $M_{j,j+1}$ and $V_{j,j+1}$ respectively. Denote the probability $p$ to be the probability moving down from $g_{j,j+1}$ to $g_{j,j}$ and $(1-p)$ to be the probability moving through the spanning branch from $g_{j,j+1}$ to $g_{j+1,j+2}$. \\

We can recall the law of total expectation which states
\begin{align}
	\E\left[ X \right] = \sum p(Y = y)\E\left[ X \mid Y = y \right].
\end{align}
If we let
\begin{align}
	X = \rnE\left[ G(t_{j+1}) \mid G(t_j) = g_{j,j+1} \right],
\end{align}
and $Y$ denotes the random variable such that
\begin{align}
	Y = \left\{ \begin{array}{cl}
		0, & \mbox{if moving to the spanning brance} \\
		1, & \mbox{otherwise}
	\end{array} \right.,
\end{align}
then
\begin{align}
	M_{j,j+1}
	& = \rnE\left[ G(t_{j+1}) \mid G(t_{j}) = g_{j,j+1} \right] \notag \\
	& = p\rnE\left[ G(t_{j}) \mid G(t_j) = g_{j,j} \right] + (1-p)\rnE\left[ G(t_{j+1},j+2) \mid G(t_{j}) = g_{j,j+1} \right] \notag \\
	& = pM_{j,j} + (1-p)g(t_{j+1},j+2),
\end{align}
which shows
\begin{align}
	p = \frac{g(t_{j+1},j+2) - M_{j,j+1}}{g(t_{j+1},j+2) - M_{j,j}}.
\end{align}

The task of deriving the relationship between $g(j+1,j+2)$ and $p$ from the law of total variance is more complicated. We will show the the result first, then break into each part. Recall the law of total variance which states
\begin{align}
	\Var\left( X \right) = \E\left[ \Var(X \mid Y) \right] + \Var\left( \E[X \mid Y] \right).
\end{align} 
Similarly we let 
\begin{align}
	X = \rnE\left[ G(t_{j+1}) \mid G(t_j) \right],
\end{align}
and $Y$ denotes the random variable such that
\begin{align}
	Y = \left\{ \begin{array}{cl}
		0, & \mbox{if moving to spanning brance} \\
		1, & \mbox{otherwise}
	\end{array} \right..
\end{align}
If the following statement is true:
\begin{align}
\label{App_B_var}
	\rnVar\left( \rnE\left[ G(t_{j+1}) \mid G(t_j) = g_{j,j+1} \right] \right)
	& = \rnE\left[ \rnVar\left( \rnE\left[ G(t_{j+1}) \mid G(t_j) = g_{j,j+1} \right] \mid Y \right) \right] \notag \\
	& + \rnVar\left( \rnE\left[ \rnE\left[ G(t_{j+1}) \mid G(t_j) = g_{j,j+1} \right] \mid Y \right] \right) \notag \\
	& = pV_{j} + p(M_{j,j} - M_{j,j+1})^2 + (1-p)(g(j,j+1) - M_{j,j+1})^2, 
\end{align}
then we have
\begin{align}
	p = \frac{V_j - (g(j,j+1) - M_{j,j+1})^2}{V_{j} + (M_{j,j} - M_{j,j+1})^2 - (g(j,j+1) - M_{j,j+1})^2}.
\end{align}

Examining the first term, $pV_j$, on the right-hand-side of equation (\ref{App_B_var}), 
\begin{align}
	\rnVar\left( \rnE\left[ G(t_{j+1}) \mid G(t_j) = g_{j,j+1} \right] \mid Y = 0 \right) = \rnVar\left( g(j+1,j+2) \right) = 0
\end{align}
since there is only one choice moving from $g_{j,j+1}$ to $g_{j+1,j+2}$. On the other hand, 
\begin{align}
	\rnVar\left( \rnE\left[ G(t_{j+1}) \mid G(t_j) = g_{j,j+1} \right] \mid Y = 1 \right) = V_{j},
\end{align}
and we know this value recursively. So
\begin{align}
	\rnE\left[ \rnVar\left( \rnE\left[ G(t_{j+1}) \mid G(t_j) = g_{j,j+1} \right] \mid Y \right) \right]
	& = (1-p)\rnVar\left( \rnE\left[ G(t_{j+1}) \mid G(t_j) = g_{j,j+1} \right] \mid Y = 0 \right) \notag \\
	& + p\rnVar\left( \rnE\left[ G(t_{j+1}) \mid G(t_j) = g_{j,j+1} \right] \mid Y = 1 \right) \notag \\
	& = p\rnVar\left( \rnE\left[ G(t_{j+1}) \mid G(t_j) = g_{j,j+1} \right] \mid Y = 1 \right) \notag \\
	& = pV_j.
\end{align}
Next, the second and third terms. Since
\begin{align}
	\rnE\left[ \rnE\left[ \rnE\left[ G(t_{j+1}) \mid G(t_j) = G(t_j,j+1) \right] \mid Y \right] \right] = \rnE\left[ G(t_{j+1}) \mid G(t_j) = G(t_j,j+1) \right] = M_{j,j+1},
\end{align}
we have
\begin{align}
	& \; \rnVar\left( \rnE\left[ \rnE\left[ G(t_{j+1}) \mid G(t_j) = g_{j,j+1} \right] \mid Y \right] \right) \notag \\
	& = \; p\left( \rnE\left[ \rnE\left[ G(t_{j+1}) \mid G(t_j) = g_{j,j+1} \right] \mid Y = 1 \right] - M_{j,j+1} \right)^2 \notag \\
	& + \; (1-p) \left( \rnE\left[ \rnE\left[ G(t_{j+1}) \mid G(t_j) = g_{j,j+1} \right] \mid Y = 0 \right] - M_{j,j+1} \right)^2 \notag \\
	& = \; p\left( \rnE\left[ G(t_{j+1}) \mid G(t_j) = g_{j,j} \right] - M_{j,j+1} \right)^2 + (1-p)\left( g(j+1,j+2) - M_{j,j+1} \right)^2 \notag \\
	& = \; p\left( M_{j,j} - M_{j,j+1} \right)^2 + (1-p)\left( g(j+1,j+2) - M_{j,j+1} \right)^2.
\end{align}

Now we have two equations and two unknown $p$ and $g(t_{j+1}, j+2)$ in the following
\begin{align}
	\left\{ \begin{array}{cl}
		p & = \frac{g(t_{j+1}, j+2) - M_{j,j+1}}{g(t_{j+1}, j+2) - M_{j,j}} \\
		p & = \frac{V_j - (g(j+1,j+2) - M_{j,j+1})^2}{V_{j} + (M_{j,j} - M_{j,j+1})^2 - (g(j+1,j+2) - M_{j,j+1})^2}
	\end{array} \right..
\end{align}
However, $p$ must be a number between $0$ and $1$. And we now show that the equations indeed yield a solution such that $p \in [0,1]$. First, the case where $M_{j,j+1} \geq M_{j,j}$ and write
\begin{align}
	f_1(x) = \frac{x - M_{j,j+1}}{x - M_{j,j}}
\end{align} 
and
\begin{align}
	f_2(x) = \frac{V_j - (x - M_{j,j+1})^2}{V_{j} + (M_{j,j} - M_{j,j+1})^2 - (x - M_{j,j+1})^2}.
\end{align}
Since $M_{j,j+1} \geq M_{j,j}$, for any $x \geq M_{j,j+1}$
\begin{align}
	0 \leq \frac{x - M_{j,j+1}}{x - M_{j,j}} \leq 1
\end{align}
and $f_1(x)$ is continuous and monotonically increasing. On the other hand, for any $x \in [M_{j,j+1}, M_{j,j+1}+\sqrt{V_j}]$, 
\begin{align}
	0 \leq \frac{V_j - (x - M_{j,j+1})^2}{V_{j} + (M_{j,j} - M_{j,j+1})^2 - (x - M_{j,j+1})^2} \leq 1
\end{align}
and $f_2(x)$ is a continuous and monotonically decreasing function. Since 
\begin{align}
	f_1(M_{j,j+1}) = 0, \qquad \qquad f_1(M_{j,j+1}+\sqrt{V_j}) > 0
\end{align}
and
\begin{align}
	f_2(M_{j,j+1}) > 0, \qquad \qquad f_x(M_{j,j+1}+\sqrt{V_j}) = 0,
\end{align}
we know that there must exist a unique $\hat{x} \in [M_{j,j+1}, M_{j,j+1}+\sqrt{V_j}]$ such that
\begin{align}
	f_1(\hat{x}) = f_x(\hat{x}) = p \in [0,1]. 
\end{align}
Alternatively, the proof is similar for the case when $M_{j,j+1} < M_{j,j}$ except the solution exists in $[M_{j,j+1}-\sqrt{V_j},M_{j,j+1}]$. The uniqueness and existence of the solution $p$ and $g(t_{j+1},j+2)$ help us solve the equations fast.

\section{Application: AA Rated Callable Corporate Bonds}
A callable bond is a bullet bond with an embedded American-style issuer call option. The option gives the issuer the right to refinance (when interest rates are low, for example), therefore the price of a callable bond is lower than the price of a bullet bond with the same coupon rate and maturity. To price the option, we can assume the underlying bullet bond price process follows a log-normal process. However, the method does not work if we simply want to price the AA corporate callable bond with its option feature. A better solution is to model the bond yield term structure instead of the individual bond price. Further discussions can be found in \cite{l2012-paradigm}. \\

The purpose of this application is to demonstrate how to simulate AA bond yield curve by the recombination trees constructed according to the algorithm described in previous sections. Particularly we consider only AA corporate bond universe, and model the curve under the \textit{stochastic-splines-model framework} in \cite{l2012-paradigm}. The model is constructed by assuming that the coefficient processes follow the exponential Ornstein-Uhlenbeck processes under the real-world measure, which later can be transformed to a risk-adjusted measure by imposing proper no-arbitrage conditions. Once the trees are built under both measures, we can simulate the bond yield curve under the real-world measure to price bullet corporate bonds, and simulate the bond yield curve under a risk-adjusted measure to price bond embedded options. Then callable bond price will be the bullet bond price minus the option value. \\

\begin{figure}[tp]
\captionsetup{labelformat=graphs}
\centering
\includegraphics[width=0.6\textwidth]{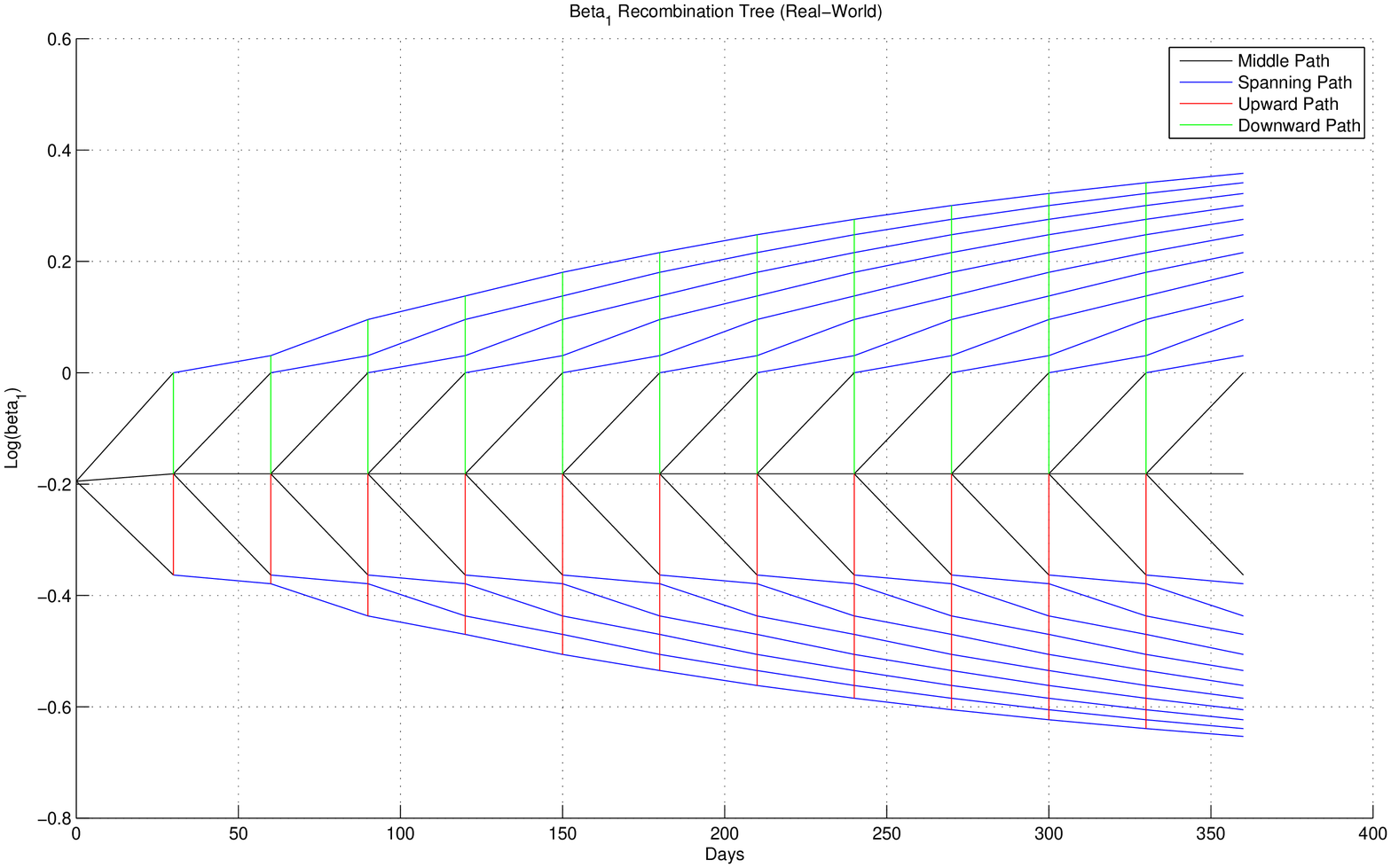}
\includegraphics[width=0.6\textwidth]{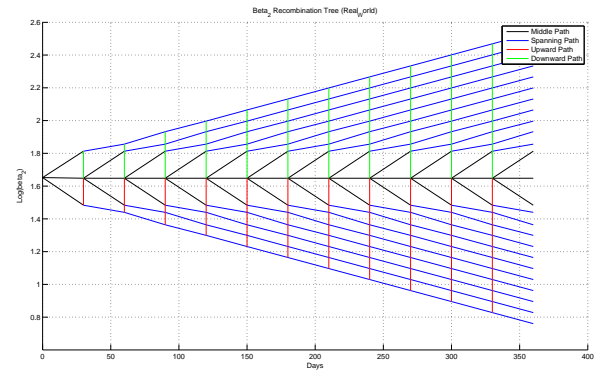}
\includegraphics[width=0.6\textwidth]{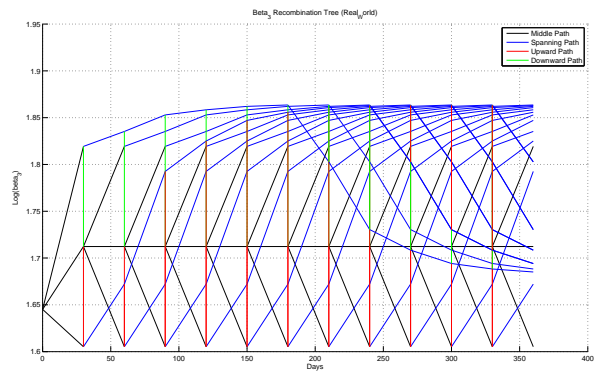}
\includegraphics[width=0.6\textwidth]{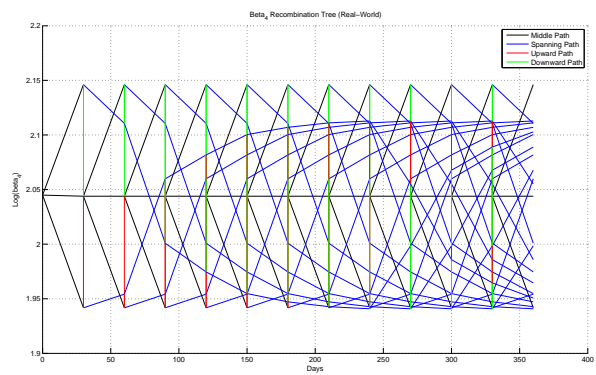}
\caption{Real-World Recombination Tree Results based on 08/31/2010 AA Corporate Bond Yield Curve Dynamics}
\label{tree_rw}
\end{figure}

\begin{figure}[tp]
\captionsetup{labelformat=graphs}
\centering
\includegraphics[width=0.6\textwidth]{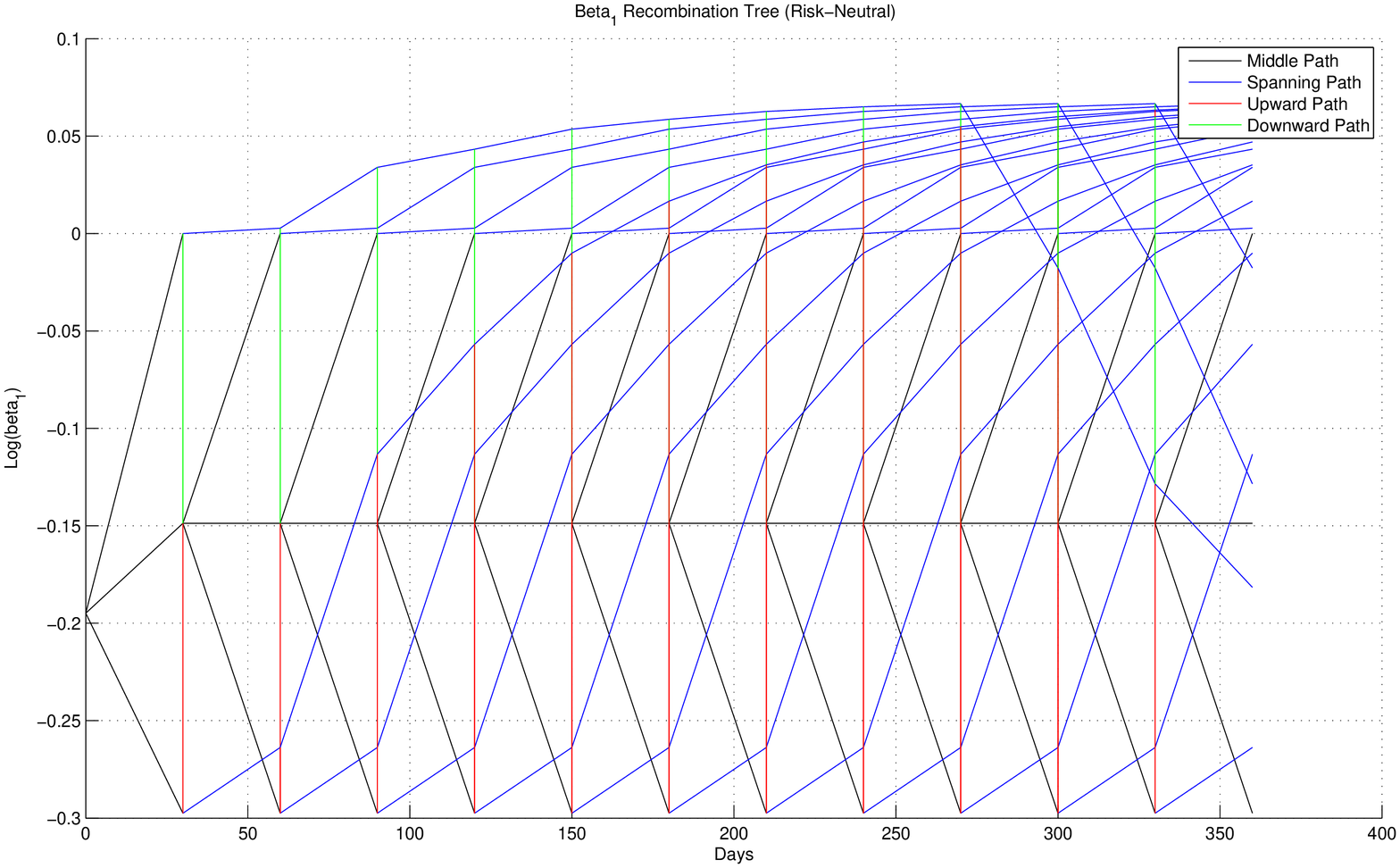}
\includegraphics[width=0.6\textwidth]{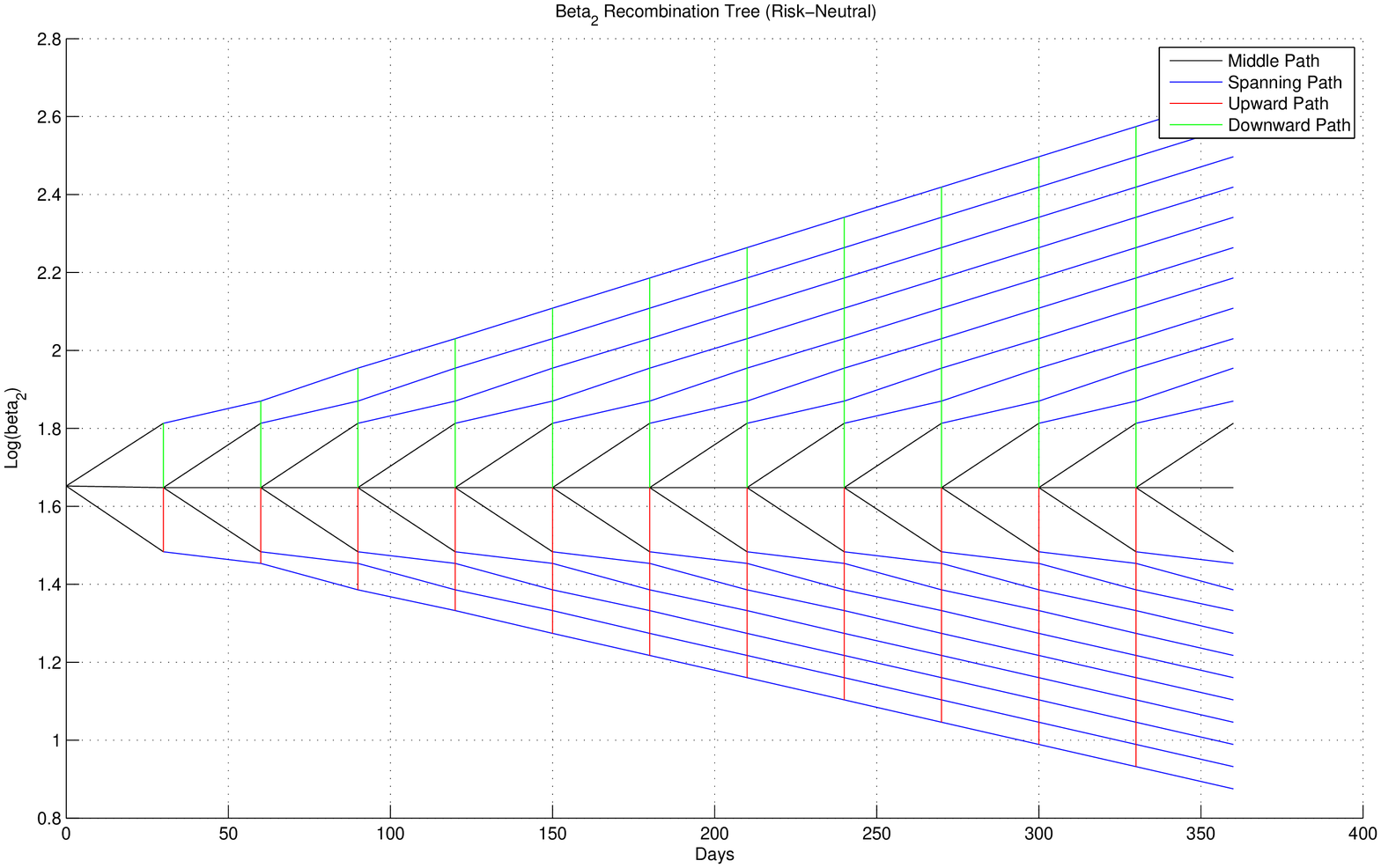}
\includegraphics[width=0.6\textwidth]{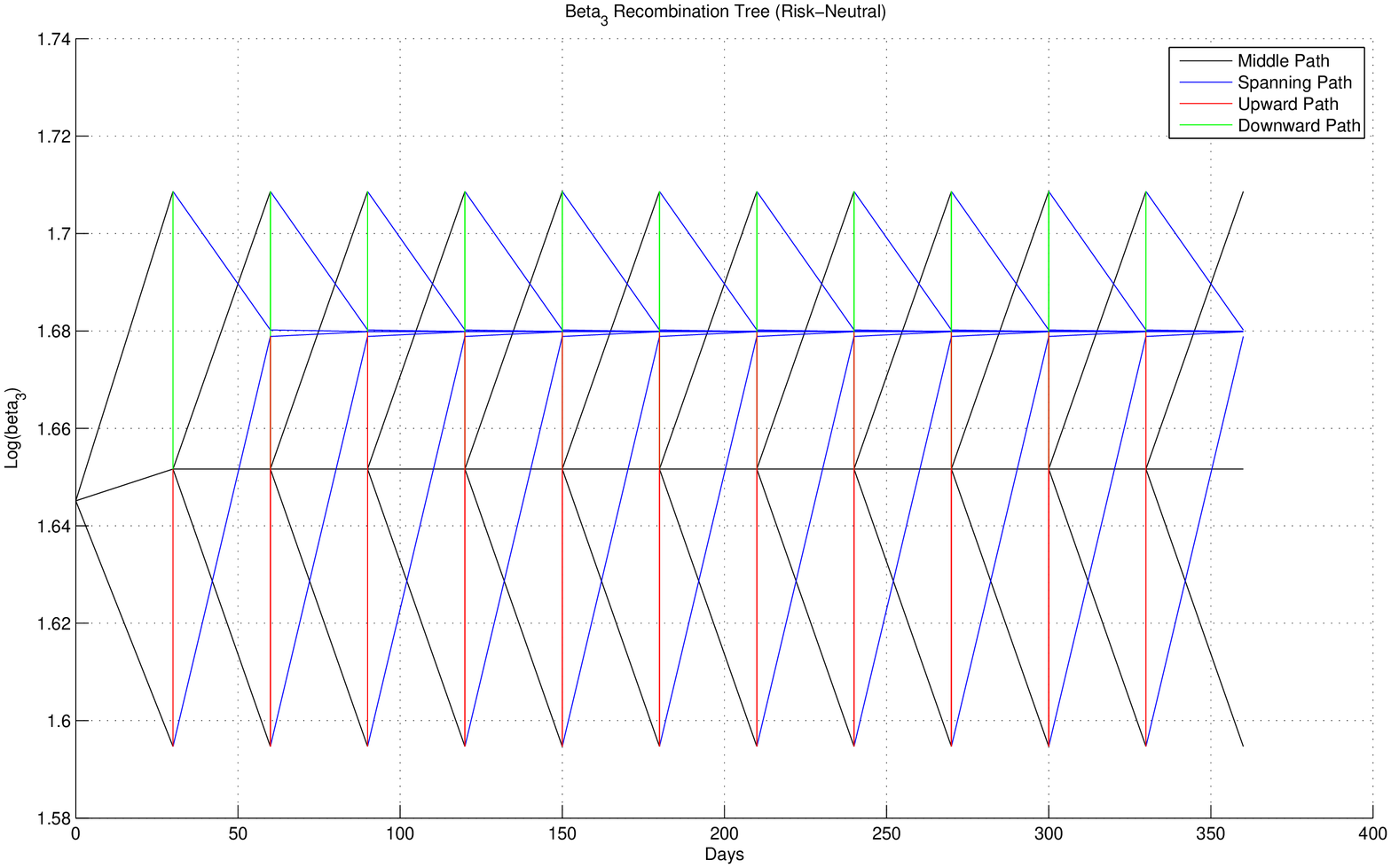}
\includegraphics[width=0.6\textwidth]{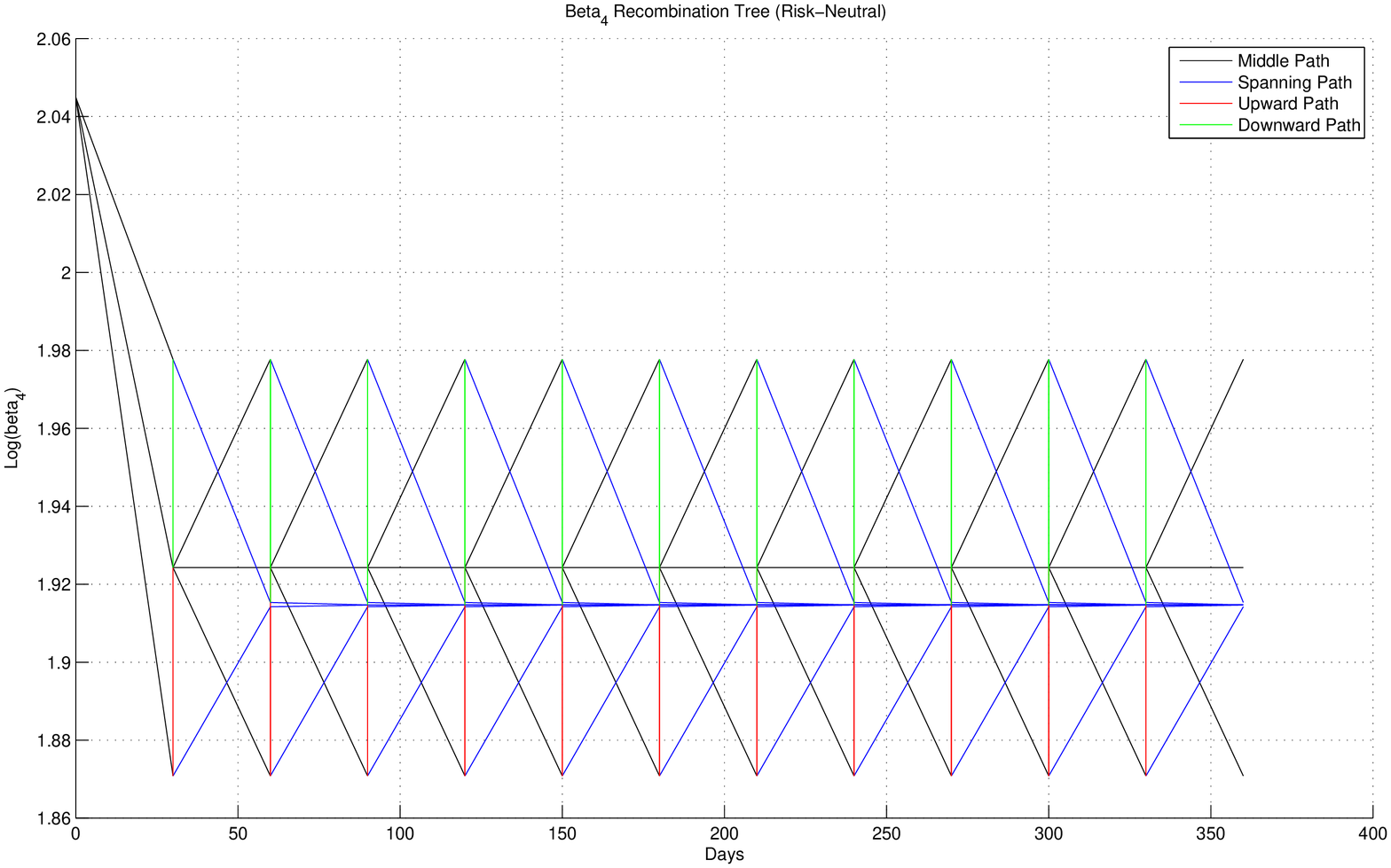}
\caption{Risk-Adjusted Recombination Tree Results based on 08/31/2010 AA Corporate Bond Yield Curve Dynamics}
\label{tree_rn}
\end{figure}

In Exhibit \ref{tree_rw}, we demonstrate those recombination tree results under the real-world measure. In Exhibit \ref{tree_rn}, we demonstrate those recombination tree results under a risk-adjusted measure. The black branches represent the center paths; the blue branches represent the spanning branches; the green branches represent the sibling branches that can only moves down; the red branches represent the sibling branches that can only moves up. 

\bibliographystyle{amsplain}
\bibliography{RefThesis}

\providecommand{\bysame}{\leavevmode\hbox to3em{\hrulefill}\thinspace}
\providecommand{\MR}{\relax\ifhmode\unskip\space\fi MR }
\providecommand{\MRhref}[2]{%
  \href{http://www.ams.org/mathscinet-getitem?mr=#1}{#2}
}
\providecommand{\href}[2]{#2}
\begin{thebibliography}{1}

\bibitem{bdt1990}
F.~Black, E.~Derman, and W.~Toy, \emph{A one-factor model of interest rates and
  its application to treasury bond options}, Financial Analysts Journal
  \textbf{46} (1990), no.~1, 33--39.

\bibitem{clrs}
T.~H. Cormen, C.~E. Leiserson, R.~L. Rivest, and C.~Stein, \emph{Introduction
  to algorithms}, 3rd ed., The MIT Press, 2009.

\bibitem{hw1990}
J.~Hull and A.~White, \emph{Pricing interest-rate-derivative securities},
  Review of Financial Studies \textbf{3} (1990), no.~4, 573--592.

\bibitem{l2012-paradigm}
P.~C.-L. Lin, \emph{A paradigm shift in interest-rate modeling}, Working Paper
  (2012).

\end{thebibliography}

\end{document}